\documentclass[11pt]{article}
\usepackage{graphicx,amsmath,amsfonts}
\usepackage[left=2cm,right=2cm,bottom=2cm,includeheadfoot,a4paper]{geometry}
\usepackage[onehalfspacing]{setspace}

\usepackage[comma,sort&compress]{natbib}
\bibpunct{[}{]}{,}{n}{,}{,}

\begin{document}
%\doi{10.1080/1478643YYxxxxxxxx}
%\issn{1478-6443}
%\issnp{1478-6435}  \jvol{86} \jnum{9} \jyear{2006} \jmonth{21 March}

\markboth{Vedmedenko, Even-Dar Mandel, and Lifshitz}{Multipolar order
  on the Penrose tiling}

\title{In search of multipolar order on the Penrose tiling}

\author{E.~Y.~Vedmedenko$^{\dag}$\thanks{Corresponding author.
    Email: vedmeden@physnet.uni-hamburg.de}, Shahar Even-Dar
  Mandel$^\ddag$, and Ron Lifshitz$^{\ddag}$ \\
  $^\dagger$ Institute for Applied Physics, University of Hamburg,\\
  Jungiusstr. 11, 20355 Hamburg, Germany\\$^{\ddag}$
  Raymond and Beverly Sackler School of Physics \& Astronomy,\\ Tel Aviv
  University, Tel Aviv 69978, Israel}

\date{May 8, 2008}

\maketitle

\begin{abstract}
  Based on Monte Carlo calculations, multipolar ordering on the
  Penrose tiling, relevant for two-dimensional molecular adsorbates on
  quasicrystalline surfaces and for nanomagnetic arrays, has been
  analyzed. These initial investigations are restricted to multipolar
  rotors of rank one through four---described by spherical harmonics
  $Y_{lm}$ with $l=1\ldots4$ and restricted to $m=0$---positioned on
  the vertices of the rhombic Penrose tiling. At first sight, the
  ground states of odd-parity multipoles seem to exhibit long-range
  order, indicated by the appearance of a superstructure in the form
  of the decagonal Hexagon-Boat-Star tiling, in agreement with
  previous investigations of dipolar systems. Yet careful analysis
  establishes that long-range order is absent in all cases investigated
  here, and only short-range order exists. This result should be taken
  as a warning for any future analysis of order in either real or
  simulated arrangements of multipoles on quasiperiodic templates.
\end{abstract}

\section{Introduction}

The concept of multipole moments is one of the most prominent and
ageless mathematical constructions in physics and chemistry. Several
decades ago the determination of electric and magnetic multipoles of
neutral and polarized molecules became a vivid domain of scientific
research because of the central role of multipole tensors for studies
of intermolecular forces~\cite{Buckingham:AdvChemPhys1967}, nonlinear
optical phenomena~\cite{Shen:1984}, electrostatic
potentials~\cite{ChemAppl}, various phenomena induced by
intermolecular forces~\cite{Birnbaum:1985}, collision effects in
Nuclear Magnetic Resonance spectroscopy~\cite{Rajan:1974}, hyperfine
interactions~\cite{Schwartz:1955}, theoretical prediction of the
geometries of van der Waals molecules~\cite{Magnasko:1990} and
electron scattering~\cite{Gianturco:1986}.  These extensive studies
have demonstrated that many molecules possess sufficiently strong
electric multipole moments. Among them are polar molecules with
asymmetric charge distribution like HF, H$_2$O, FCl, HCCl$_2$,
etc. having a permanent dipole moment; neutral Ne, Ar, Kr, Xe, O$_2$,
F$_2$, D$_2$, CO$_2$ etc. possessing quadrupole moments; polyatomic
SiF$_4$, B$_4$Cl$_4$, giant Keplerate molecule Fe$_{30}$ and CF$_4$
having strong octopolar and (HSi)$_8$)O$_{12}$, (CH$_3$Si)$_8$O$_{12}$
hexadecapolar contributions. Many organic substances possess
multipolar moments as well. These are quadrupoles like benzene,
3,4,9,10-perylenetetracarboxylicdianhydride (better known as PTCDA),
cyanogen (N$\equiv$C-C$\equiv$ N), 1,1-dichloroethene,
$cis-1,2$-dichloroethene; octopoles like methane or cyanogen and more
complicated complexes having higher order
contributions~\cite{Stone:1996,VedmedenkoBOOK07}.  Many of these
molecules and molecular complexes can be adsorbed on solid surfaces.
The arrays of adsorbates interact to a large extent via classical
electrostatic multipolar interactions~\cite{VedmedenkoBOOK07} as
confirmed by Raman spectroscopy and nuclear magnetic resonance
experiments. Often the multipolar interactions lead to complex
ordering of the moments.

Quite independently, many new problems, that are not characteristic
for bulk materials, arise at the nanoscale. One of the interesting
aspects vividly discussed nowadays is the interparticle interaction in
magnetic arrays. Magnetic properties of artificially structured and
self-organized magnetic media belong to the central questions of
nanomagnetism as they give access to new phenomena that can be used in
technology~\cite{Bland,Aign:1998,Vedmedenko04}. Recently the
importance of multipolar magnetostatic contributions for magnetization
reversal in densely packed ensembles of particles has been pointed out
theoretically~\cite{Vedmedenko:PSS2007,Politi:PRB2002}. Here, the
effect of the multipolar moments is two-fold. First, it influences the
collective magnetic ordering in an array, and second, it changes the
nucleation fields due to the stabilization of magnetization near the
edges of neighboring particles.

Hence, the knowledge of multipolar phase transitions and ground states
is extremely important for a variety of applications as well as for
fundamental understanding of physics and chemistry of solid state
systems. With the recent ability to use the surfaces of real
quasicrystals as templates upon which a variety of different particles
can be adsorbed~\cite{ledieu:2006,mcgrath}, it has become timely to
study the ground states of multipoles on aperiodic substrates as
well~\cite{Thiel:PhilMag2008}.  Yet, in contrast to the rather
well-studied multipolar ground states on periodic lattices, the data
for aperiodic tilings is quite limited\cite{Vedmedenko:2006,%
  Vedmedenko:MPL05,VGW04,VGW05}, although there exists some
group-theoretical analysis~\cite{Lifshitz:RevModPhys1997,%
  Lifshitz:PRL1998,Lifshitz:MatSci2000,Lifshitz2,Lifshitz:Acta2004,%
  Lifshitz:Enc}, describing the possible allowed symmetries of
quasiperiodic multipolar arrangements, that may be used as a guide for
our study.\footnote{Some related results exist for studies of quantum
  magnetic models on quasicrystals~\cite{Grimm:1997,Matsuo00,%
    Matsuo02,Wessel,Jagannathan97,Hida01}.  See also the discussion of
  magnetism in quasicrystals in this
  issue~\cite{Hippert:PhilMag2008}.}  The aim of this work is to
initiate an extensive theoretical study of such order, beginning with
a theoretical calculation of the ground states of multipolar rotors on
the rhombic Penrose tiling.

\section{Methods}

In this study we investigate ground states of systems of multipoles,
arranged on the vertices of the rhombic Penrose tiling, by means of
Monte-Carlo simulations. In order to calculate any order of
interaction within reasonable effort, we introduce the Hamiltonian in
spherical coordinates into the conventional MC scheme, and derive the
stable low temperature configurations
\cite{Popelier:2001,Vedmedenko:2005}.  The Hamiltonian of the
multipolar interaction reads
\begin{equation}
H=\frac{1}{4 \pi \mu_0} \!\!\sum_{
        \substack{
            A\neq B\\
            l_A l_B m_A m_B
            }
        }
\!\!\!\!T_{l_A l_B m_A
m_B}(\vec{R}_{AB})Q_{l_A m_A}^A Q_{l_B m_B}^B\label{eq:ww}
\end{equation}
where $Q_{l_A m_A}^A$ and $Q_{l_B m_B}^B$ are the moments of
multipoles $A$ and $B$ expressed in spherical harmonics, where $l$ and
$m$ correspond to the standard two degrees of freedom on a sphere. The
coupling coefficient $T_{l_A l_B m_A m_B}(\vec{R}_{AB})$ is the
geometric interaction tensor
\begin{eqnarray}
  T_{l_A l_B m_A m_B}(\vec{R}_{AB}) &= &(-1)^{l_B}
  I_{l_A+l_B\,m_A+m_B}^\ast(\vec{R}_{AB})\phantom{\frac{1}{1}}\nonumber\\  
  &\times &\sqrt{
    \frac{(l_A+l_B-m_A-m_B)!}{(l_A-m_A)!(l_B-m_B)!}
    \frac{(l_A+l_B+m_A+m_B)!}{(l_A+m_A)!(l_B+m_B)!} },
\label{eq:tensor_2}
\end{eqnarray}
where the dependence on the interparticle distance vector
$\vec{R}_{AB}$, between multipoles on sites $A$ and $B$, is given by
the complex conjugate of the irregular normalized spherical harmonic
function $I_{l\,m}(\vec{r}) = \sqrt{\frac{4\pi}{2l+1}}
\frac{Y_{l\,m}(\theta,\varphi)}{r^{l+1}}$.

\begin{figure}[t]
  \centerline{\includegraphics[width=0.5\columnwidth]{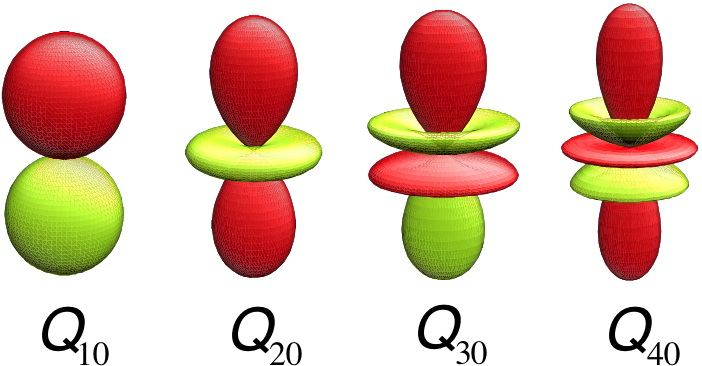}}
  {\caption{(Color online) Cylindrically symmetric multipoles $Q_{l0}$
      with $l=1...4$, represented using an equipotential surface, with
      color reflecting the sign of the charge on the internal side of
      the surface. Note that multipoles with even (odd) $l$ are
      symmetric (antisymmetric) with respect to the inversion of the
      cylindrical axis. Thus, odd-parity multipole can be thought of
      as arrows $\uparrow$, while even-parity multipoles as
      double-headed arrows $\updownarrow$.}
\label{fig:MP}}
\end{figure}

For the sake of simplicity, the Monte-Carlo simulations presented here
are restricted to cylindrically symmetric multipole moments $Q_{lm}$,
or \emph{rotors}, with $l=1...4$ and $m=0$, whereas generally $m$ can
take any value between $-l$ and $l$. These four multipolar rotors are
depicted schematically in Figure~\ref{fig:MP}, represented using 
equipotential surfaces. The restriction to $m=0$ is quite
limiting and will be relaxed in future extensions of this study,
nevertheless it is a good starting point providing interesting
results. The moments are placed on the vertices of a finite patch of
the two-dimensional rhombic Penrose tiling, using open boundary
conditions. The patches are square or rectangular in shape, containing
up to $1000$ multipole moments. We also use circular patches to verify
that our results are not affected by the shape of the sample. The
simulations are performed with an algorithm especially designed for
long-range systems: the local fields at each site are computed at the
beginning of the simulation and are only updated when a rotation
attempt is accepted \cite{Vedmedenko:2005}.  To prevent artificial
effects we do not use a cutoff in the evaluation of the multipolar
coupling.

In contrast to MC schemes for usual magnetic systems, where only
restricted rotations of the magnetic moment are often used
\cite{Nowak:PRL2000}, the rotational space is sampled continuously,
\emph{i.e.}, a moment can assume any new angle. This is especially
important in complex multipolar ensembles as these interactions might
favor large angles between neighboring spins. An extremely slow
annealing procedure with up to 150 temperature steps is applied. To
avoid metastable states we perform two different simulations of the
same system simultaneously, starting at different seeds for the random
number generator to ensure that the samples take different paths
towards equilibrium. Only when both samples reach the same stable
energy level is it taken for granted that the system has reached
its equilibrium.

\section{Ground States of Classical Multipolar Rotors on the Penrose Tiling}

The symmetry of a charge distribution around a particle determines its
non-zero multipole moments and whether moments of odd or even rank
appear. The parity of the multipole moments has a strong impact on the
ground state. We therefore consider the two cases separately below. As
can be seen in Fig.~\ref{fig:MP}, once we restrict the value of $m$ to
be 0, the resulting multipoles with even $l$ are symmetric with
respect to the inversion of the cylindrical axis, whereas those with
odd parity are antisymmetric with respect to inversion about this axis.
Consequently, odd-parity rotors can be represented geometrically as
arrows, with light-colored (yellow online) ``tails'' and dark-colored
(red online) ``heads'', that like magnetic moments tend to align
head-to-tail. On the other hand, even-parity rotors can be described
as double-headed arrows with a repulsive interaction between the heads
of neighboring rotors. In the case of the $Q_{20}$ rotors, the two
heads are attracted exactly to the central oppositely-charged regions
of nearby rotors. In the case of the $Q_{40}$ rotors the attraction is
not exactly to the center, but rather to one of the two off-center
oppositely-changed regions. As we describe below, these simple
geometric observations suffice to explain all the results that
we present here. Note that the charge flipping operation---a
color symmetry operation~\cite{Lifshitz:RevModPhys1997} that switches
between red and yellow---is equivalent to the inversion of the
cylindrical axis, mentioned above, but only for the case of odd-parity
rotors. In the case of even-parity rotors, charge flipping takes the
rotors to oppositely charged counterparts that we do not consider in
our simulations here. In the future it would be interesting to see
what happens if one introduces charge flipping of individual moments
as an additional MC step for the even-parity multipoles.

Experimental equivalents of the charge distributions having odd rank
multipolar contributions include uniformly polarized magnetic and
ferroelectric nanoparticles \cite{Dzyaloshinskii:1998,Mikuszeit:2004,%
  Scheinfein:1996,Warin,Politi:PRB2002}.  Generally, such particles
may possess a mix of dipolar $Q_{1m}$, octopolar $Q_{3m}$,
dotriacontapolar $Q_{5m}$ and possibly even higher-order
contributions.  However, for certain geometries some of the multipole
moments may become extinct.  For example, a tetragonal prism with
equal height, width, and length---which is therefore a
cube---possesses strong dipolar but a zero octopole moment, while its
strongly elongated or very flat counterparts have strong octopolar
contributions. The dependence of the strength of multipole moments on
the effective aspect ratio and shape of a particle can be found in
\cite{Mikuszeit:2004,Mikuszeit:2005}. To the experimental systems
possessing multipoles of even order belong molecular adsorbates
including H$_2$, N$_2$, CO on salts (e.g.  boron nitride) or metal
surfaces, organic PTCDA molecules on Ag, and methane on graphite.

\subsection{Odd-Parity Multipole Moments: Dipoles and Octopoles}

\begin{figure}[tb]
  \begin{center}
    \includegraphics[width=0.48\columnwidth]{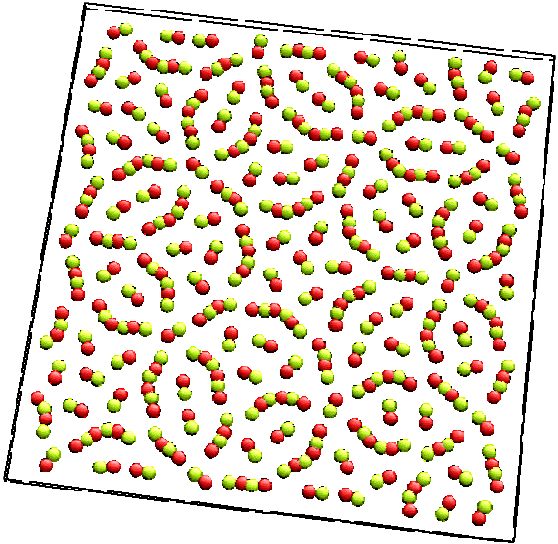}
    \includegraphics[width=0.48\columnwidth]{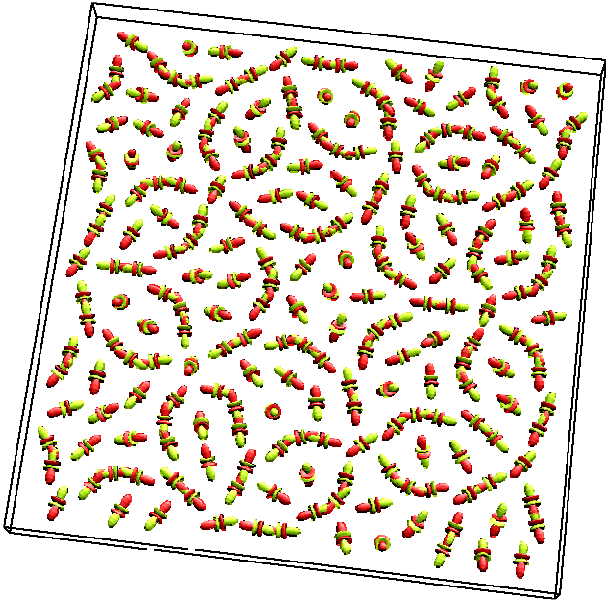}
  \end{center}
  \caption{(Color online) Ground-state configurations of odd-parity multipolar
    rotors. (Left) dipole moments; (Right) octopole moments. These
    configurations possess short-range order, owing to the strong
    head-to-tail attraction of neighboring multipoles, which is
    sufficient to highlight the decagonal Hexagon-Boat-Star Tiling.
    \label{fig:DiOcto}}
\end{figure}

In previous theoretical studies of dipolar ordering (multipole moments
of rank one $Q_{1m}$) on the Penrose tiling, performed in Cartesian
coordinates~\cite{Vedmedenko:2003,EMatVedmedenko,%
  Vedmedenko:MPL05}, a decagonal pattern with long-range order was
proposed as the ground state. Here, after careful and extensive
analysis, we find clear evidence for short-range order, with very
interesting geometric properties, yet we see no evidence for the
emergence of long-range multipolar order.  The ground-state
configurations of odd-parity multipoles are shown in
Fig.~\ref{fig:DiOcto}. At first sight these configurations seem to
possess very nice long-range multipolar order, as one clearly sees a
superstructure in the form of the familiar decagonal Hexagon-Boat-Star
(HBS) tiling~\cite{Cockayne:PRL1998}. Decagonal rings are clearly
visible in both cases, each subdivided into a single boat and a pair
of hexagonal tiles. Between the decagons one easily identifies the
star-shaped tiles. This aesthetic arrangement of the multipoles may
lead one to the incorrect conclusion that there exists long-range
order of the multipoles on the underlying Penrose tiling. Yet careful
analysis shows that this arrangement stems from the short-range
head-to-tail attraction of neighboring multipoles and exhibits
\emph{no long-range order}. We would like to use this example as a
warning for any future analysis of either real or simulated
arrangements of multipoles on quasiperiodic templates. One has to be
very careful in the analysis of order in such quasiperiodic
structures, as it is harder to visually comprehend them in real space
than their periodic counterparts.

As it turns out, the HBS tiling is simply outlined by pairs and
triplets of multipoles that are separated by the short diagonals of
the thin ($36^\circ$) rhombic tiles of the Penrose tiling. This
separation, which is the shortest interparticle separation on the
Penrose tiling, sets the largest energy scale in the system. As such,
these pair and triplet chains are the first to order as the
temperature is lowered. Because their positions and orientations are
strictly inherited from the Penrose tiling, their ordering on the
short scale suffices to outline the HBS tiling that one clearly
observes. The existence of short-range order in the orientation of the
multipoles is verified quantitatively through a statistical analysis.
The absolute orientation of the multipoles, projected onto the plane,
is clearly peaked along the 20 directions ($n\pi/10$ for $n=1\ldots
20$), dictated by the Penrose tiling. The dipolar histogram is
more strongly peaked relative to the octopolar one owing to the fact
that the octopolar ground state possesses, on average, a larger
out-of-plane component.  A frequency distribution of the angle between
nearest neighboring moments is also peaked at the characteristic
angles inherited from the relative orientation of thin rhombic tiles
on the Penrose tiling, and is significantly less-pronounced for the
octopolar moments due to their substantial out-of plane protrusion. We
note that other than the fact that the octopolar arrangement contains
a larger average out-of-plane component, and a slightly less-perfect
short-range order within the plane, the two cases are quite
similar. The difference stems from the fact that there is some amount
of attraction of the arrow heads to the central oppositely-charged
regions of neighboring octopoles (see Fig.~\ref{fig:MP}).

We note that in magnetic systems the strength of the multipolar
interactions can be tuned by the shape of the particles, or their size
relative to the interparticle separations. The octopolar contribution
may become very large for $R_{AB}<s$, with $s$ being the lateral size
of a particle. The dipolar contributions are sizable for
$R_{AB}<5s-10s$. For very small interparticle separations the
decagonal structure might become disordered due to the octopolar
contributions, while for very large separations disorder may appear
because of the weakness of the dipolar coupling.  This implies that
there exists a critical separation $R_{AB}^c$ for which the
short-range ordering of odd-parity multipoles on the Penrose tiling is
maximal. For typical particle shapes used in
experiments~\cite{Vedmedenko:2005} this critical distance is of order
of $1s-2s$.

The multipoles that lie within the HBS tiles are disordered, as can be
verified by simple inspection. Nevertheless, one could still imagine a
situation in which the multipoles that lie on the edges of the HBS
tiles are long-range ordered while the internal multipoles are
not. Yet upon further inspection one finds that multipoles that lie on
the edges of the HBS tiles are disordered as well, as their direction
changes randomly from one pair or triplet chain to the next. This
disorder is a direct consequence of the frustration that arises
whenever the ends of three such chains meet together. This can be
seen, for example, at the 5 vertices of the central star tile in both
configurations, shown in Fig.~\ref{fig:DiOcto}. Our observation of the
lack of long-range order is confirmed quantitatively by performing a
Fourier analysis of the ground state configurations. By examining the
different components of the multipolar fields, as well as various
functions of the components, we can say with certainly that such order
is lacking, as the calculated Fourier spectra show no additional Bragg
peaks when compared to the Fourier spectrum of the tiling
itself. Thus, the only long-range order that is observed is in the
positions of the multipoles, inherited from the Penrose tiling, and
not from their relative orientation. To be sure, we have also
calculated the Fourier spectrum of a randomly oriented configuration
of multipole moments on the vertices of the Penrose tiling , created
using a random number generator. The outcome strongly resembles those
of the ground-state configurations.
 
A natural question to ask at this point is whether the lack of
long-range order of our arrow-like objects on the Penrose tiling is a
geometrical property of the tiling. If we disregard the physics, or
the energetics, would it be possible to find a geometrical arrangement
of ordered arrows on the Penrose tiling? The positive answer to this
question was given by one of us many years ago~\cite{L95}, where two
such configurations---in which the arrows are located at the tile
centers, rather than their vertices---were demonstrated and analyzed
using the tools of color~\cite{Lifshitz:RevModPhys1997} and
magnetic~\cite{Lifshitz:PRL1998,Lifshitz:Enc} symmetry. If this is the
case, then one should wonder whether our lack of order is a physical
consequence of the particular type of interaction
Hamiltonian~(\ref{eq:ww}) that we use. Would it be possible to arrange
arrow-like objects on the Penrose tiling with sufficiently weak
frustration, or possibly no frustration at all, and obtain an ordered
ground state of some other Hamiltonian? It turns out that the answer
to this question is also yes, as was demonstrated almost a decade ago
by Cockayne and Widom~\cite{Cockayne:PRL1998}. In their case the
arrows are positioned on the edges of the Penrose tiles and represent
the chemical ordering of pairs of Cu and Co atoms in a model of a
decagonal Al-Cu-Co quasicrystal. Thus, if one imagines the yellow- and
red-colored circles of our dipole (Fig.~\ref{fig:MP}, left) as being
Copper and Cobalt atoms, one can obtain the desired ground state using
the model of Cockayne and Widom. It remains to be seen whether an
appropriate modification of our interaction Hamiltonian, can lead to
an ordered ground state of multipoles.

\subsection{Even-Parity Multipole Moments: Quadrupoles and
  Hexadecapoles}

\begin{figure}[tb]
  \begin{center}
    \includegraphics[width=0.48\columnwidth]{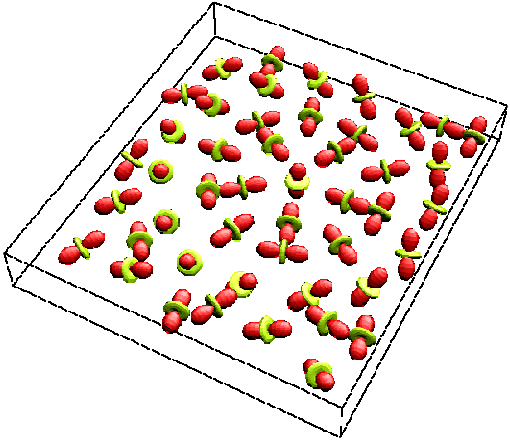}
    \includegraphics[width=0.48\columnwidth]{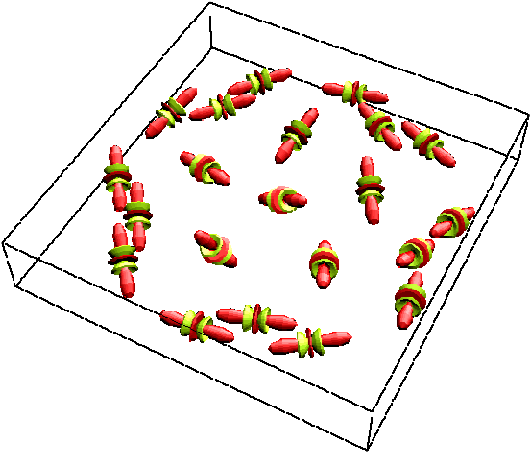}
  \end{center}
  \caption{(Color online) Ground-state configurations of even-parity multipolar
    rotors. (Left) quadrupole moments; (Right) hexadecapole moments. These
    configurations possess some degree of short-range order which is
    harder to establish than in the case of odd-parity multipoles of
    Fig.~\ref{fig:DiOcto}. One clearly sees that the red-colored edges
    are attracted to the central, or slightly off-centered, yellow
    regions of neighboring multipoles. This causes the quadrupoles to
    align at exactly $90^\circ$, taking advantage of out-of-plane
    orientations. In the case of hexadecapoles the local arrangements
    are less trivial, although they clearly exhibit local symmetry.
    \label{fig:QH}}
\end{figure}

Patches taken from the ground-state configurations of even-parity
multipoles are shown in Fig.~\ref{fig:QH}. These configurations do not
possess any long-range order, as confirmed by careful Fourier
analysis. Short-range order is clearly present, but it is not as easy
to analyze and quantify as the case of odd-order multipoles where it
shows up in the form of the HBS superstructure. Nevertheless, one can
see that the quadrupole moments tend to align predominantly at
$90^\circ$ angles, as well as $72^\circ$, owing to the strong
attraction of the edges of one quadrupole to the center of its
neighboring quadrupole. In this manner the quadrupoles can form
nicely-ordered local decagonal structures, like the one seen at the
center of Fig.~\ref{fig:QH} (Left). The local arrangements of the
hexadecapoles are even more complicated to describe, with the edges of
one attracting the opposite charges located slightly off the centers
of its neighbors. Nevertheless, one can see nice triplet chains of
hexadecapoles, and an overall nicely-ordered local pentagonal
configuration like the one shown in Fig.~\ref{fig:QH} (Right). As
mentioned above, this order does not extend beyond the short range.

\section{Summary}

In conclusion, we have studied the low-temperature stable multipolar
ground states on the Penrose tiling by theoretical means. We have shown
that long-range order is absent in all cases investigated here.
Nevertheless, short-range order exists owing to the strong interaction
between particles of closest separation---those that are separated by
the short diagonals of the thin Penrose tiles.  In the case of
odd-parity multipoles, this short-range order, combined with the
underlying structure of the rhombic Penrose tiling, suffices to
outline a superstructure in the form of the decagonal HBS tiling. The
multipoles lack long-range order despite the appearance of the HBS
superstructure, because the orientations of the moments on the edges
of the HBS tiles are disordered due to 3-body frustration.

Further investigations are clearly necessary in order to seek out
other possibilities of long-range multipolar order on the Penrose
tiling. These should relax the $m=0$ restriction imposed here,
consider alternative forms of interparticle interaction, as well
as explore other possibilities for the positions of particles on
the quasiperiodic surface, such as those corresponding to the
so-called ``dark-stars'' on the 5-fold surfaces of real icosahedral
quasicrystals~\cite{mcgrath,Thiel:PhilMag2008,ledieu:2006,Schaub:1994}. 

%%%%%%%%%%%%%%%%%%%%%%%%%%%%%%%%%%%%
 
\section*{Acknowledgments} 

Financial support from the Deutsche Forschungsgemeinschaft in the
framework of the part project A11 of the SFB 668 is gratefully
acknowledged. This research is also supported by the Israel Science
Foundation through Grant 684/06.

\bibliographystyle{unsrt}
\bibliography{EYV-biblio_12_2007}

\label{lastpage}

\end{document}